# Software Agents Interaction Algorithms in Virtual Learning Environment

Dr. Zahi A.M. Abu Sarhan
Faculty of Information Technology: Software Engineering Department
Applied Science Private University
Amman, Jordan

*Abstract*— This paper highlights the multi-agent learning virtual environment and agent's communication algorithms. The researcher proposed three algorithms required software agent's interaction in virtual learning information system environment. The first proposed algorithm is agents interaction localization algorithm, the second one is the dynamic agents distribution algorithm (load distribution algorithm), and the third model is Agent communication algorithm based on using agents intermediaries. The main objectives of these algorithms are to reduce the response time for any agents' changes in virtual learning environment (VLE) by increasing the information exchange intensity between software agents and reduce the overall network load, and to improve the communication between mobile agents in distributed information system to support effectiveness. Finally the paper describe the algorithms of information exchange between mobile agents in VLE based on the expansion of the address structure and the use of an agent, intermediary agents, matchmaking agents ,brokers and their entrepreneurial functions.

## I. INTRODUCTION

At present, most of information system in various fields based on software agents technology, these systems can implemented as distributed information systems as centralized information systems, and one of these complex systems is distance or electronic learning systems. For the e-learning systems wildly used centralized technology development and information systems operation. The situation is still likes the 90's of the last century, when there were a large number of scattered, technically and semantically mixed databases, used locally. Different kinds of attempts to create a central system for utilization all information system resources to achieve valuable functionality and flexibility have yielded little success. The wildly used today's Internet technologies and software agents techniques with web access, solves the problem partially. In this way, there is a distributed access to information and the information resources and by the way it keeps it to be centralized, which makes them not always sufficiently relevant, high redundancy, technological, and semantic diversity [1].

In this context, it is urgent the task to think about creating a fully decentralized peer information systems to support e-learning process, enabling flexible integration into virtual and logical platform intelligently accomplish requested task and usefully invoking all available information resources, thus creating a unified virtual learning platform for fully efficient interaction of all learning subjects. Effective technology for implementation distributed information systems of this class is the mobile software agents' technology [1, 2]. These systems must not only distribute access to information, but also decentralized data storage and processing, and solving problems related with information resources semantic diversity.

The developed multi-agent system implements a virtual learning environment (VLE), in which real learning processes related with development and implementation of all comprehensive forms of academic technological innovation in learning and education and any comprehensive ideas that can be implemented as relevant information processes[4]. The subjects of learning can be presented as software agents that interact with each other in a single VLE and perform the interest of its owners, forming an open multi-agent system with a decentralized architecture.

Software agent is computer system, which is found in some environment and is capable of autonomous action in this environment in order to meet its design objectives [4]. Software agents have characteristics that make them suitable for complex functions. Such features include: autonomy, interaction, reactivity, activity, intelligence and mobility [5, 6].

This paper presents the software agents interaction models and the algorithms for learning activities in the virtual learning information systems environment, enabling faster response to changes in the agents virtual environment simulating innovative field by increasing the intensity of information exchange between the agents and to reduce the overall load on the network. Indicating ways to enhance the interaction among mobile agents in an apportioned multi-agent system of information to back up learners and teachers action.

## II. PROBLEM STATEMENT

Multi-agent systems (MAS) technology - a new pattern of information technology, focused on the sharing of scientific and technological achievements and benefits which provide ideas and methods of artificial intelligence (AI), the current local and global computer networks, distributed databases and distributed computing, hardware and software tools to support the theory of distribution and transparency.

Relevance of distributed AI and MAS, in accordance with the works [7, 9], determined by the complexity of diversity modern organizational and technical systems, complexity, and tasks distributions, large volumes of Information flow and information processing time. Agent-oriented approach is widely used in various fields that required solutions for complex distributed tasks such as combined product design, re-engineering information processes and the construction of virtual systems, and e-commerce systems, distributed computer programs development.

The greatest complexity in the theoretical studies and practical modern MAS implementations are issues related to the agents' interaction processes in the collective solving problems with high practical complexity and relevance, as each agent solves specific subtask, has only a partial idea about the general task and should continuously interact with other agents. Therefore, at present the most relevant in the MAS theory and technology is the issues related with creating intelligent agents interaction models in the MAS.

In inter-host software agents interaction in a virtual learning environment, arise a set of issues related with increasing the network load, and reducing the information exchange intensity between agents which depends on the speed and bandwidth of the network connections. Increasing the request delivery time and the needs time to find the correct agent for interaction. As a suitable solution for these problems, proposed a method based on the decomposition a common information space in which agents operate in virtual platform (platform represents some separate network host) and move intensively interacting agents to these platforms in order to combining agents in coalition. The proposed solution can be implemented in two mutually complementary algorithms software agents' interaction: the algorithm of inter-host software agent's interaction (Transformation the inter-host agent's interaction to intra-host interaction) and software agent's dynamic allocation algorithm (balancing load between system hosts) [10].

However, the architecture design of decentralized open information systems, arise problem related with determining the location of dynamically distributed mobile agents, that are moving between network hosts over time, as well as minimizing the data losses possibility and the transmitting messages between agents delay time. The proposed solution in this paper based on the expansion of the agent address structure and the information about its current location in the network and local caching this information on the system hosts, which will allow sending messages directly from the sender agent to the recipient agent, as well as provide the ability to search agents with the joint action through intermediary agents, using their brokering services [11] and matchmaking services [12] functions. Algorithm searching agents' initiators executed in the agents intermediaries address space.

## III. MULTI-AGENT VIRTUAL LEARNING ENVIRONMENT

The modern virtual learning environment (VLE) is a software system designed to support teaching and learning. VLEs generally function on the World Wide Web, and, therefore, they can be utilized both on and off-campus, provided that the users are authorized and can access the Internet [12]. This surmount over the restriction of traditional face-to-face learning, and guarantees that learning is neither limited to location nor time. VLEs can contains complex of interactions between teachers, learners and learning contents, by adding agents and environment to this structure, the VLE can be presented as a complex of agents interaction and every agent represents the interest of the learning structure components, such as learner agent, teacher agent, content agent, and every agent interacts with other agents to accomplish requested task. However the set of interacted agents can contain "learner –learner interaction agent, learner-teacher interaction agent, learner-content interaction agent, teacher-teacher interaction agent, teacher-content interaction agent, content-content interaction agent, learner-environment interaction agent, teacher-environment interaction agent, content-environment interaction agent", VLEs become more popular and included in many college organizations all over the world. It is not only because of their versatility, but also because they provide an extensive range of tools or features, such as content distribution, evaluation, emails[14, 15].

Based on the learning process components the representation of VLE can be presented as a set of agents that interact together, shown in figure 1.

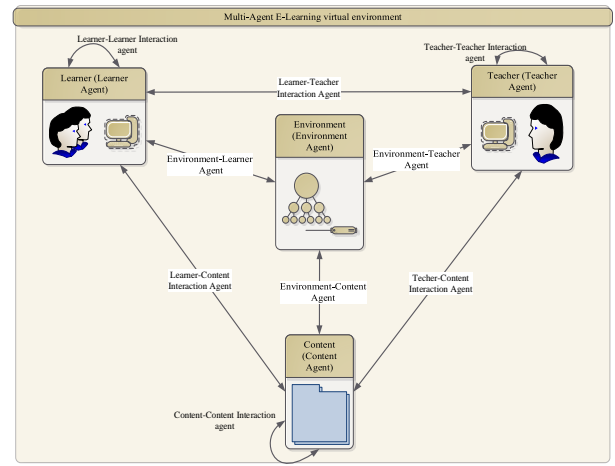

Figure 1. Virtual learning environment based on multi-agents architecture.

## IV. AGENTS COMMUNICATION MODEL

Agents Communication Model in an open multi-agent information system that supports e-learning process, according to the work [14], can be classified by the interaction character in the following categories: direct (point-to-point communication), interaction and moderating (throw intermediaries or server agents) interaction. In point to point agents' model communication message delivered directly from agent sender to the agent receiver, even if the agent receiver is a mobile agent (a message can be sent through a set of computer network host's). In the intermediary agent model communication mainly uses agents brokering services or matchmaking services agent's intermediaries, which allow agents to communicate with each other, this model provides multipoint and anonymous agent's interaction. These two types of services have their advantages and disadvantages: brokering services are more efficient, while matchmaking services are more flexible [10].

Depending on the agent-receiver location in the network agent's communication model, according to [6, 9], may be also classified as follows: inter-host communication and intra-host communication. When the agent-sender and agent-receiver located on the same network host, the messages transmission between these both agents called intra-host communication. When they were located on two different network hosts, the messages transmission between agents delivered via the network nods, and this type of communication called the inter-host communication. Even if the Internet data rate has greatly increased, inter-host agent's communication still would take greater time than interaction with intra-host communication. Therefore, it is advisable to reform agent's inter-host communication to intra-host communication, which would reduce the time required for messages transmission and increase the interaction intensity in the agent's communication process.

The main disadvantages of presented agent's interaction models can be listed as follows:

– At inter-host interaction: the network over load increases, and the information exchange intensity between agents reduced due to increasing message delivery times, agents reaction for dynamically changing events of environment or the absence of any response to change the environment in message losing case;

– At the agents' migration between system hosts there was a problem of the needed responder agent location determining for interaction during any time moment, which in its turn results to increasing in messages loss probability during their transfer and reducing the effectiveness of synchronized agent's interaction.

## V. AGENT-BASED ALGORITHMS VIRTUAL LEARNING PLATFORMS FORMATION

The higher effect obtained from using the virtual learning environment (VLE), based on the biggest of its internal volume (the number of registered requests), Agents representing the Learning subjects interests, and placement of information databases hosts [17]. However, the naturally growth of the system volume leads to increase the information elements search task complexity, and the selection of learning structures options due polynomial growth of alternatives quantity. In order that the system no longer runs under its own unrestricted growth, needed some self-organization algorithms, allowing dynamically reorganize its internal structure to reduce the amount of processed and transmitted over the communication lines data during the learning offerings placement and retrieval on the media hubs, and forming a potentially effective learning structures.

Self-organization is the automatic generation within-VLE virtual learning platforms (VLP), combining agents with similar interests in groups. Formation the VLP based on a the register distribution support method for peer-to-peer hosts with the implicit treelike organization [9] in which as an organizing structure tree uses the hierarchical learning domain model. Learning platforms formation carried out by displaying the agent's purposes on the treelike conceptual domain model, Subsequent localization the main part of search and other agents requests inside the group and further analysis the communications activity with each other. The interests similarity leads to the fact that the most active and informative agents communication focused inside learning platform, whereas outside platform the information exchange is less active, thus an exchange object generalized (smaller by volume) agents learning offers or requests [18].

This approach allows not only reducing the total communication amount between agents, but also due to the use of mobile agents to transform inter-host in intra-host communication. This, in turn, reduces the network load in case of the distributed system implementation. The main agent based algorithms that enhance e-leaning systems efficiency are agent interactions localization algorithm within hosts by forming groups actively communicates agents (virtual learning platforms) and dynamic transfer load redistribution by implementing agents group migration algorithm. These algorithms provide ultimately, the conversion agents inter host interactions in intra-host. In order to convert the inter-host interactions in intra host interactions, the system agents can be distributed the network hosts depending on their interaction nature and behavior. Since the agents interaction nature can be continuously changed, the agents distribution on the network hosts should be dynamic. In this case, the agents automatically distributed on multiple network hosts, depending on their interaction nature, some hosts may be overwhelmed by the several migrating agents to them. Therefore, the proposed agents dynamic distribution algorithm, that based on their interaction behavior should be complemented by another agents distribution algorithm which support uniform load distribution between network hosts.

In the proposed multi-agent system [7], each host equipped with an agent-based platform (agent representative), which is a agents local execution environment, in which software agents operate and interact with each other. Implemented agents interaction algorithms involve a sequence of similar operating phases and use common components that are present on each agent platform.

To arrange a dynamic agents distribution on each agent system platform the following components should operate. The Message Manager (MM), coordinating the messages exchange between the system agents; and the System Monitor (SM), periodically checking the load on the current network host; and the Agent Distribution Manager (ADM), is responsible for the agents dynamic distribution and analysis the models of agents interaction behavior and character; and finally the Agent Migration Manager (AMM), displacing agents on other platforms (hosts) and managing agents migration between network hosts.

*A. Agents interaction Localization algorithm*

The agent's interaction localization algorithm will contain four phases as shown in figure 2.

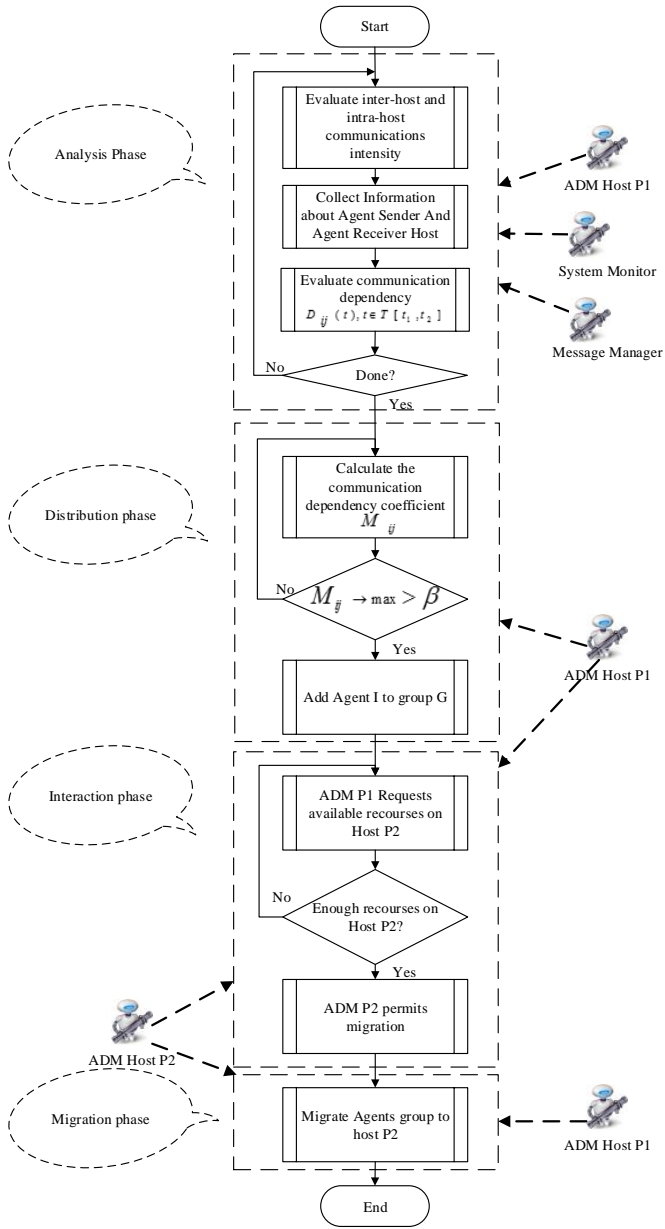

Figure 2. Agent's interaction localization algorithm

- Analysis phase. The Agent's Distribution Manager (ADM) evaluates inter-host and intra-host system agents' communications intensity, involving in this stage system monitor and messages manager. ADM also uses all information as about agent-sender, as about agent-receiver host. ADM periodically evaluates communication dependency $D_{ij}(t)$ at time $t \in T = [t_1, t_2]$ between the agent $i$ and agents host $j$ as follows:

$$D_{ij}(t) = d \left( \frac{R_{ij}(t)}{\sum_k R_{ik}(t)} \right) + (1-d) D_{ij}(t-1) \quad (1)$$

Where $D_{ij}(t)$ - the number of messages sent by the agent $i$ to host $j$ agents over the time period $T[t_1, t_2]$, $\delta$ - is the coefficient, which characterizes the relative dependency of new information in relation to the outdated (expired) and used to ignore the temporary intensive interaction with the agents in particular agent platform; $D_{ij}(t-1)$ - is defined as the value of the same communication dependency in the previous time interval.

- Agents Distribution Phase. After a specified number of analysis stage repetitions ADM calculates the communication dependency coefficient between the current host agent $n$ and all other hosts in the system. Communication dependency coefficient $M_{ij}$ between agent $i$ and host $j$ agents can be given by:

$$M_{ij} = \left( \frac{D_{ij}}{D_{in}} \right), j \neq n \quad (2)$$

When the maximum value of the communication dependency coefficient over a predefined edge $\beta$, ADM current host includes under consideration the agent in a group of agents, located on a remote system host:

$$B = \arg_j \max(M_{ij}) \wedge (M_{ik} > \beta) \rightarrow \varphi_i \in G_k \quad (3)$$

Where $\varphi_i$ is an agent $i$, $G_k$ indicates the agents group $k$, and under $\arg_j$ hereinafter understood operation that returns the value of $j$, where the ratio of communication dependence coefficient $M_{ij}$ takes the maximum value.

- Interaction Phase. Before moving the selected agents group from the host $P_1$ to the receiver host $P_2$ ADM host $P_1$ interacts with ADM host $P_2$. ADM $P_2$ checks the current memory host state, CPU utilization, and the number of agents hosted in this host, by using system monitor. If the host $P_2$ has enough free system resources for new agents, ADM host $P_2$ authorizes the agents group migration from host $P_1$ to host $P_2$.

- Agents' Migration Phase. Accomplishing the agents' selection operation for migration and receiving positive response from ADM receiver host, the sender host ADM jointly with Agents Migration Manager (AMM) initiates the selected agents' group migration to the receiver host.

Agents' interaction localization algorithm assumes the analysis of dynamic changes in interaction patterns (models) between agents, but this algorithm can overloaded some system hosts due to the large number of agents which can be moved to the hosts.

When a host is overloaded, the system monitor detects this state and activates the agents' redistribution process, which is not only based on individual agent movement between network hosts, but the whole interacting agents group intensively interacts with each other.

### B. The agents dynamic distribution algorithm

The agents dynamic distribution algorithm (load distribution algorithm) among system hosts consists five phases: Analysis phase agents grouping phase, agent groups distribution phase, agents interaction phase and the agents migration phase.

- Analysis Phase. Each system host monitor periodically checks the agent platform state in which agents interacting on this host. And by calling special system functions collects information about the host physical resources such as (current CPU load and free memory space). When the system monitor defines that the host is overloaded, it will activate the agents' distribution procedure between system hosts. As a host overload criterion can be offered the maximum number of agents operated on the host.

When ADM gets notification about congestion from the system monitor, it starts the local agents interaction monitor procedure, preliminary in order to locally splitting interacted and cooperated agents into groups.

The group formed from agents that have most similar interests and goals, which are intensively, interact with each other by exchanging messages. Information about agents and agents groups registered in the special register. The information analysis provided on the register, allows to evaluate the load on the system hosts and determine the intensity between agents and intergroup and inter-host communication levels, which, in turn, allows to select loaded and unloaded hosts in the system, and implement the dynamic agents redistribution and agents groups among the system hosts, in other words the agents or agents groups movement with certain characteristics from overloaded hosts to less loaded hosts in the system, containing similar potential agents interests. After a predetermining the time interval ADM updates communication dependences between the agents and the locally formed agents groups on different network hosts [10].

Communication dependence of $D_{ij}(t)$ between agent $i$ and agents group $j$ at the moment $t$ can be determined by the formula:

$$D_{ij}(t) = \delta\left(\frac{R_{ij}(t)}{\sum_K R_{iK}(t)}\right) + (1-\delta) D_{ij}(t-1) \quad (4)$$

Where $D_{ij}(t)$ - is the number of messages sent by the agent $i$ to the agents group $j$ - over the time period $T[t_1, t_2]$, $\delta$ - is the coefficient, which characterizes the relative importance of new information in relation to the outdated (expired) and used. In this case the expression $\sum_K R_{iK}(t)$ shows the number of messages sent by agent $i$ to any agent jointly functioning on the common host.

- Agents Grouping Phase. After accomplishing a number of monitoring phases repetitions each $i$ agent is overridden in local agents group with new index $j^*$ and given by:

$$j^* = arg_j max(M_{ij}(t)) \rightarrow \varphi_i \in A_{j^*} \quad (5)$$

Where $A_{j^*}$ indicates $j^*$ - local group agents.

Monitoring and agents grouping phases are repeated several times. After each agents grouping phase information about local communication dependencies between agents zeroed.

- Agent Groups Distribution Phase. After performing a certain number repetitions of monitoring and agents grouping phases, ADM depending on agent platform status takes the decision to move the agents group to other system host. The selection of moved agents based on the communication dependence between agent groups and the system hosts.

Communication Dependence $W_{ij}$ between $i$ - agents group and the $j$ system host based on the summation of the communication dependence between all members of agents group and the system host:

$$W_{ij}(t) = \sum_{k \in A_i} C_{kj}(t) \quad (6)$$

Where $A_i$ - all agents indexes set, belonging to $i$ - agents group, and $C_{kj}(t)$ - communication dependence between agent $k$ and $j$ system host in the current time $t$. The agents group $i^*$, which have a smallest dependence on the current host and can be selected by the following rule:

$$i^* = arg_i max\left(\sum_{j, j \neq n} \frac{W_{ij}}{W_{in}}\right) \quad (7)$$

Where $n$ is the current host (Agent platform) number [10].

The receiving agent platform $j^*$, selected by $i$ agents group determined using communication dependence between $i$ agents group and $j^*$ system host as follows:

$$j^* = arg_j max(W_{ij}), j \neq n \quad (8)$$

Where $n$ is the current host (Agent platform) number.

- Agents Interaction Phase. When in the network defined the receiving host for agent groups, the sender host ADM begins interaction with the corresponding receiving host ADM. If the receiving ADM allows agents movement, sending host ADM starts the agents migrating phase. Otherwise, the sending host ADM polls other system hosts ADM as long as it finds a suitable host for a scheduled agents migration (movement). If no host is unable to take the agents group, the interaction phase fails and after a certain time period the agents distribution (allocation) algorithm on the sending host restarted, and the process repeats again.

Agents interaction phase of this algorithm is similar to the agents interaction localization algorithm. Yet, the interaction level between these two algorithms is very different: agents' interaction localization algorithm occurs at the agent level, whereas the algorithm load distribution between the system hosts implemented at the agent groups level. If the agent receiver-platform representation has sufficient system resources for all agents belong to the selected local agents group, ADM agent receiver-platform representation may authorize the agents' group movement. Otherwise, the system receiver host rejects the request to move the selected agents group; the host is unable to accept only part of agents group.

- Agents Migration Phase. When ADM sender host receives a response with acceptance to move the agents group from ADM receiver host, ADM sender host initiates migration the selected local agents group to the receiver host.

Each agent-based representation independently performs its own agents distribution algorithm in accordance with the available information about the load on the system host, where it is located, and the agents interaction character functioning in agents platform [10]. The agents dynamic distribution algorithm shown in figure 3.

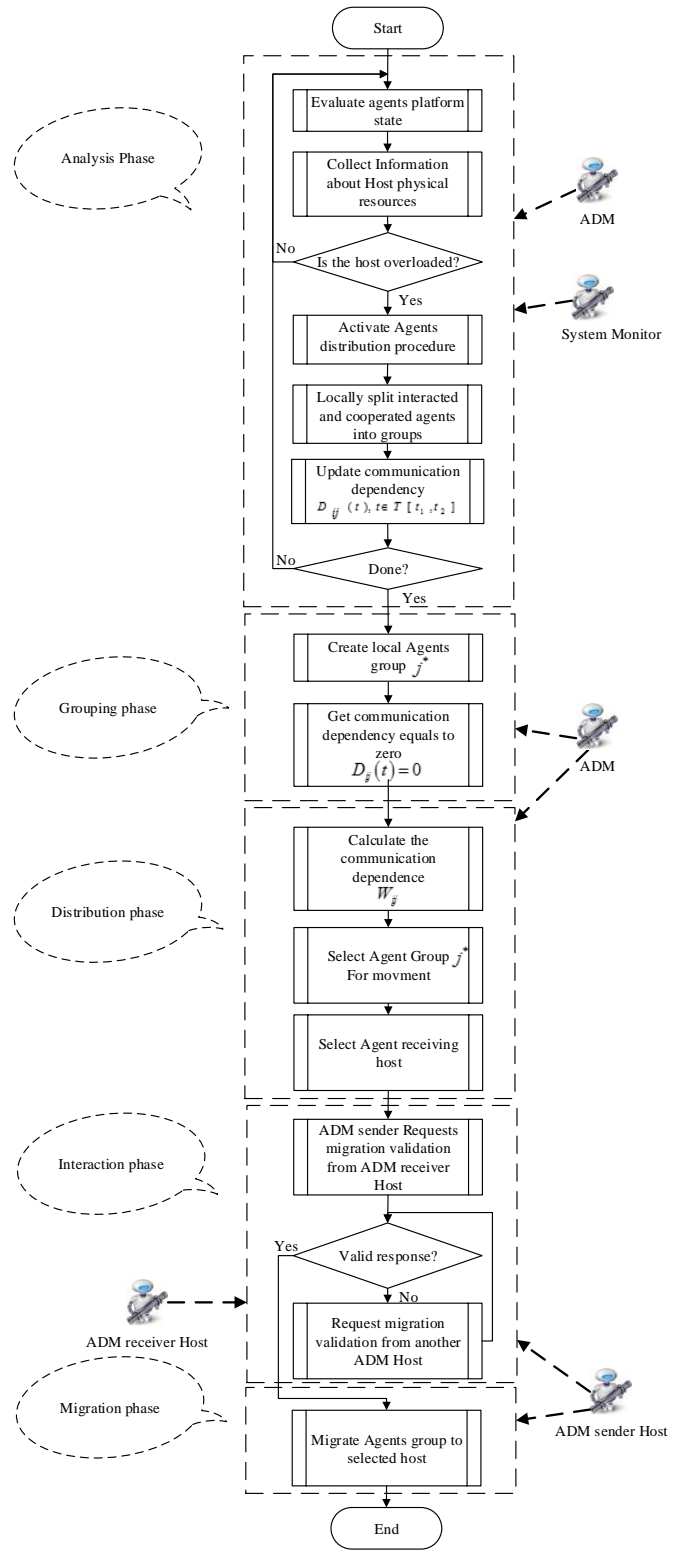

Figure 3. Agent's dynamic distribution algorithm

### C. Agent communication model based on using agents intermediary

In open multi-agent systems with extremely dynamic character (in any time may come in or/and go out new agents,

and be connect new hosts), each individual agent cannot have full information about all other agents in the same information and communication environment. In this environment, a joint operation of network services agents intermediaries (brokerage and matchmaking services) are highly effective in finding potential agents for interaction [19, 22]. Software agents can find the other agents names using matchmaking services or send messages to other agents using brokering services, using their attributes, such as methods, operation modes, features or nicknames instead of their real names. To register the agents names in distributed virtual learning environment can be used the dedicated server for agents' names. Agent Intermediary functions implemented as a separate component in its architecture, or they can be implemented as an independent specialized application called agents intermediaries [24].

Interaction mediating between the agents includes brokers' agents or matchmakers' agents. Brokers forward the messages given by the agent sender to the network hosts where the agent receiver located, whereas matchmakers' only provides the agents intending to send messages, all information about the agent receiver location, in other words the final message was delivered to the agent receiver brokers, while the matchmakers only help sender agents to deliver messages to the agent receiver, providing information about their locations. Considering the number of steps within transmitting massages, it can be concluded that the brokerage services are more effective than matchmakers, since brokerage services typically require two steps to transmit massage, while matchmakers requires three steps.

Using shared memory area, that called data fragments space, which controlled by intermediary agents, agents can register their attributes together with the names in this area, and they can communicate each other by using the attributes information about other agents, this information can be extracted from this area. The basis of most developed intermediary agents is Linda-model [9, 10]. The proposed model presented in figure 4.

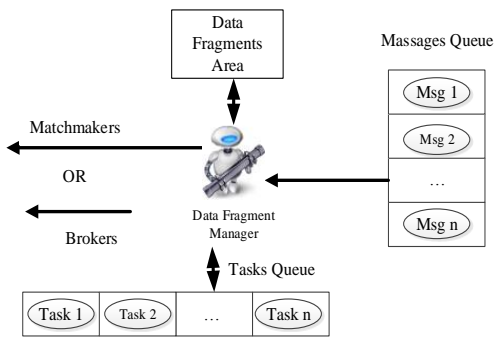

Figure 4. Agent communication model based on using agents intermediary

The model allows software agents to use intermediary agents, and the provided services by them to search agents with the similar interests and provides the basis method of using intermediary agents and their functions. The intermediaries agents don't only control the data fragments, but can be also used as own agents search algorithms "joint activity" and algorithms provided by agents' initiators. This is the main difference between the proposed method of using intermediary agents and their features compared to the existing ones.

VI. CONCLUSION

This paper discusses separated agent-based algorithms for Virtual learning environment information system effective functioning and distributed hosts interaction algorithms. Described various implementation approaches of agents communication algorithms. Analyzed the existent software agents communication models. Identified weaknesses of these models, based on that proposed approaches to improve the mobile agents cooperation in distributed Virtual learning environments.

Based on the modification of existing agents interaction models, developed algorithms for data exchange among mobile software agents that increase the agents efficiency for distributed data processing and reduce the total transmitted data amount over the network:

- Agent communication algorithm based on the using agents brokers and their functions;

However, developed software agents interaction algorithms for information system activities in virtual learning environments, that allowing faster agents response for environment changes in which an agent functioning, by increasing information exchange intensity between agents and reduce the overall load on the network, those algorithms are Agents interaction localization algorithm transformation inter-host communication to interact as intra-host communication; and agents dynamic distribution algorithm (balancing load between the system hosts).

Based on developed software agents interaction models and algorithms, proposed the minimizing inter-host communication in problem-oriented distributed systems method. The method is based on software agents classification in semantic space, presented as conceptual domain model, and agents inter-host communication transformation in intra-host. Implementation this method provides a reduction in communication infrastructure load and improving availability coefficient for software agents application services.

ACKNOWLEDGEMENT

The author is grateful to the Applied Science Private University, Amman, Jordan, for the financial support granted to cover the publication fee of this research article

REFERENCES

[1] A. Маслобоев, Мультиагентная технология формирования виртуальных бизнес-площадок в едином информационно-коммуникационном пространстве развития инноваций, Publisher, City, 2009.
[2] B.M. Balachandran, M. Enkhsaikhan, Developing multi-agent e-commerce applications with JADE, in: Knowledge-Based Intelligent Information and Engineering Systems, Springer, 2007, pp. 941-949.
[3] D. Tavangarian, M.E. Leypold, K. Nölting, M. Röser, D. Voigt, Is e-learning the Solution for Individual Learning, Publisher, City, 2004.

[4] T. Erl, A. Karmarkar, P. Walmsley, H. Haas, L.U. Yalcinalp, K. Liu, D. Orchard, A. Tost, J. Pasley, Web service contract design and versioning for SOA, Prentice Hall, 2009.

[5] F.P. Brooks Jr, The design of design: Essays from a computer scientist, Pearson Education, 2010.

[6] E. Platon, N. Sabouret, S. Honiden, Tag Interactions in MultiAgent Systems: Environment Support, Publisher, City, 2005.

[7] S. Alouf, F. Huet, P. Nain, Forwarders vs. centralized server: an evaluation of two approaches for locating mobile agents, Publisher, City, 2002.

[8] N. Carriero, D. Gelernter, Linda in context, Publisher, City, 1989.

[9] B. Chen, H.H. Cheng, J. Palen, Integrating mobile agent technology with multi-agent systems for distributed traffic detection and management systems, Publisher, City, 2009.

[10] А.В. Маслобоев, Модели и алгоритмы взаимодействия программных агентов в виртуальной бизнес-среде развития инноваций, Publisher, City, 2009.

[11] K. Malik, Use of knowledge brokering services in the innovation process, in: Management of Innovation and Technology (ICMIT), 2012 IEEE International Conference on, IEEE, 2012, pp. 273-278.

[12] T.A. Farrag, A.I. Saleh, H.A. Ali, Semantic web services matchmaking: Semantic distance-based approach, Publisher, City, 2013.

[13] Z. Xu, Z. Yin, A. El Saddik, A Web Services Oriented Framework for Dynamic E-Learning Systems, Publisher, City.

[14] A. Targamadze, R. Petrauskiene, Classification of distance learning agents, Publisher, City, 2010.

[15] E.M. Van Raaij, J.J. Schepers, The acceptance and use of a virtual learning environment in China, Publisher, City, 2008.

[16] A. Di Stefano, C. Santoro, Locating mobile agents in a wide distributed environment, Publisher, City, 2002.

[17] W.F. McComas, Virtual Learning Environment, in: The Language of Science Education, Springer, 2014, pp. 110-110.

[18] E. Sangineto, An Adaptive E-Learning Platform for Personalized Course Generation, Publisher, City, 2008.

[19] S.L. Greenspan, E. Hadar, Using cloud brokering services for an opportunistic cloud offering, in, Google Patents, 2013.

[20] S. Muthaiyah, L. Kerschberg, Brokering Web Services via a Hybrid Ontology Mediation Approach Using Multi Agent Systems (MAS), Publisher, City, 2010.

[21] P. Khanna, B. Babu, Cloud Computing Brokering Service: A Trust Framework, in: CLOUD COMPUTING 2012, The Third International Conference on Cloud Computing, GRIDs, and Virtualization, 2012, pp. 206-212.

[22] J.-b. LI, C.-h. LI, Research on the Development of the Cooperation among Governments, Enterprises, Universities, Research Institutes, Financial Units and Intermediary Agents [J], Publisher, City, 2010.

## AUTHORS PROFILE

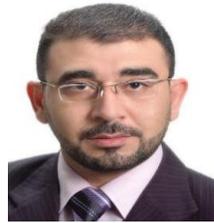

**Dr. Zahi A. M. Abu Sarhan** Received the M.S. and PhD degrees in Computerized Control Automated Systems and Progressive Information Technologies from Kharkov National University of Radio Electronics, Kharkov in 1998 and 2004, respectively. During 2004-2008, I was an Assistant Professor at the Economics and Administrative science/ MIS Department at Applied Science University. Since 2008, I am an Assistant Professor at the Faculty of Information technology, Applied Science University in Jordan. Research interests include: Information system reengineering, Service oriented architecture, software agents, agents theory, agents behavior.